\documentclass[twocolumn,traditabstract]{aa}
\usepackage{txfonts}
\usepackage[dvips]{graphics}
\usepackage[dvips]{graphicx}
\usepackage{graphicx}
\usepackage{longtable}
\usepackage{color}
\usepackage{url}
\usepackage{dcolumn} 

\usepackage{natbib}

\newcommand\PKS{PKS~2155-304}

\bibpunct{(}{)}{;}{a}{}{,} 

\begin{document}

\title{The minijets-in-a-jet statistical model and the RMS-flux correlation}
\author{J. Biteau and B. Giebels}
\institute{Laboratoire Leprince-Ringuet, Ecole Polytechnique, CNRS/IN2P3, F-91128 Palaiseau, France}
\authorrunning{Biteau \& Giebels}
\titlerunning{The minijet-in-a-jet statistical model}
\date{Received 19/07/2012 / Accepted 12/10/2012}

\abstract{

The flux variability of blazars at very high energies does not have a clear origin. Flux variations on time scales down to the minute suggest that variability originates in the jet, where a relativistic boost can shorten the observed time scale, while the linear relation between the flux and its RMS or the skewness of the flux distribution suggests that the variability stems from multiplicative processes, which are associated in some models with the accretion disk. We study the RMS-flux relation and emphasize its link to Pareto distributions, characterized by a power-law probability density function. Such distributions are naturally generated within a minijets-in-a-jet statistical model, in which boosted emitting regions are isotropically oriented within the bulk relativistic flow of a jet. We prove that, within this model, the flux of a single minijet is proportional to its RMS. This relation still holds when considering a large number of emitting regions, for which the distribution of the total flux is skewed and could be interpreted as being log-normal. The minijets-in-a-jet statistical model reconciles the fast variations and the statistical properties of the flux of blazars at very high energies.}

\offprints{biteau(at)in2p3.fr, berrie(at)in2p3.fr}
\maketitle

\section{Introduction}

Active galactic nuclei (AGN) are astrophysical sources emitting broad-band electromagnetic radiation that can be observed from radio wavelengths to very high energies (VHE, $E \gtrsim 100$~GeV). The current framework for understanding the emission from these objects depicts them as composed of a disk, feeding a super massive black hole (SMBH), possibly with a jet on each side of the accretion plane. The orientation of the system with regard to the observer's line of sight is one of the key parameters for unifying the various subclasses of AGN \citep{Urry}. The class of blazars for which the jet is probably directed within a few degrees toward the observer is of particular interest for the study of non-thermal processes occurring in astrophysical relativistic flow.

The VHE emission of blazars, such as the BL Lac object Mrk~421 or the flat spectrum radio quasar PKS~1222+21, can sometimes exhibit doubling times of the order of ten minutes \citep{1996Natur.383..319G,2011ApJ...730L...8A} while variability on the minute scale is observed in light curves of BL Lac objects, such as Mrk~501 \citep{2007ApJ...669..862A} or PKS~2155--304 \citep{2007ApJ...664L..71A}. The high quality of the VHE data on the latter object has enabled the characterization of various statistical properties of the emission, such as a skewed flux distribution, interpreted as a log-normal distribution, and a linear relation between the sample RMS of the flux and the sample flux \citep{2155_2010}. 

These properties were initially studied in the X-ray emission of the galactic black hole binary Cygnus X-1 \citep{2001MNRAS.323L..26U,UttMcVa} and are now observed in other accreting objects, such as the blazar BL Lac \citep{2009A&A...503..797G}, or non-aligned AGNs, such as the Seyfert galaxies NGC~4051 \citep{2004MNRAS.348..783M} and IRAS~13224-3809 \citep{1538-4357-612-1-L21}. 

They are sometimes interpreted as arising from multiplicative processes, originating e.g. in the accretion disk. The product of random variables whose logarithms have finite moments is indeed asymptotically log-normal. Furthermore, flux log-normality implies a linear relation between the flux and its RMS \citep[see Appendix~D of][]{UttMcVa}. 

The statistical properties of non-thermal light curves are key ingredients for answering the crucial question of the origin of variability in AGNs: do the variations of the flux come from the jet itself, i.e. the medium that conveys the accelerated particles, or from the disk that could modulate the jet emission. In accretion models such as developed by \citet{1997MNRAS.292..679L}, inward propagating fluctuations in the disk from different radii cumulate in a multiplicative way \citep[see e.g.][]{2006MNRAS.367..801A}, which could explain the log-normality and the linear RMS-flux relation. 

However, variability on time scales shorter than the light-crossing time $T_G~=~2GM/c^3$ are difficult to achieve \citep{2012MNRAS.420..604N} in disks. Transient emission on time scales $t_{\rm var}~\sim~T_G / 50$ are easier to locate in a jet with a bulk Lorentz factor $\Gamma$ \citep[unless generated by a companion system, as e.g. in][]{2010A&A...520A..23R}. Phenomena occurring on time scales $T_G$ then appear at $t_{\rm var}~\sim~T_G / \delta$ to the observer, where $\delta$ is the Doppler factor of the jet\footnote{The link between the Doppler factor and the bulk Lorentz factor is discussed extensively in the following.}.

Values of the Doppler factor $\delta \geq 50$ are not uncommon for VHE blazars and permit both minute-time-scale variations \citep{2007ApJ...664L..71A} and an optical thinness to the emitted VHE $\gamma$-rays \citep{2008MNRAS.384L..19B}. \citet{2008MNRAS.386L..28G} propose a needle-in-a-jet model, where fast minijets closely aligned with the line of sight could explain the variations. \citet{2009MNRAS.395L..29G} have investigated a jets-in-a-jet scenario, involving reconnection and a Poynting-dominated flow. \citet{2012MNRAS.420..604N} considered a turbulent process, modelled by a random motion of minijets inside the jet. Models involving minijets have also been invoked to explain the fast variations in the emission of radio galaxies \citep{2010MNRAS.402.1649G}, though the time scales of variations observed at VHE are much longer than for blazars, approximately a day for M~87 \citep{2012ApJ...746..151A}.

These models are additive (the flux is given by the sum of the contributions of several regions) and, provided there are finite moments for the individual components, the  central limit theorem (CLT) can be applied. But the Gaussian flux distribution expected from the CLT is not skewed, as is the observed one.\\
\par

We seek a way out the jet-or-disk origin dilemma by studying the implications of the observed linearity between the sample RMS and the flux (Sect.~\ref{UsualExp}). We discuss the properties of Pareto-distributed fluxes in Sect.~\ref{Pareto} and discuss the relevance of the CLT for these random variables. We show in Sect.~\ref{Sec:Model} that Pareto distributions are naturally generated by the minijets-in-a-jet statistical model and discuss the consequences of this model in Sect.~\ref{Sec:CCL}.

\section{Additive processes and Pareto distributions}

Modelling of VHE blazar variability should reproduce the short time scale of the variations and the statistical properties of the flux. While the stochastic nature of the flux is understood if the emission is modulated by the disk of the SMBH, the observed time scales are difficult to explain. Inversely, a jet origin of the variability is often rejected because the models are additive and should not be able to reproduce the skewness of the flux nor the linear RMS-flux relation. This argument is studied in the next section.

\subsection{Additive and multiplicative processes}\label{UsualExp}

Within additive models, the observed flux is the sum of the contributions of several (and potentially many) regions. If the components are modelled as independent, identically distributed random variables (hereafter iid) with finite moments, their sum should follow a normal distribution according to the CLT, assuming that the number of regions is large enough - typically more than a few tens. A natural outcome of such a process would then be symmetric flux distributions, in disagreement with the high-flux tails frequently observed in light curves. Moreover, the addition of incoherent variations in various components should result in a flux being independent of its variance. A translation into statistical terms would be that the sample mean\footnote{i.e. averaged on short time periods in the light curve.} and the sample RMS of a distribution are independent if and only if the underlying distribution is a Gaussian \citep[see the discussion in the Appendix~D of][]{UttMcVa}.

{\it A contrario}, multiplicative processes naturally generate log-normality, with a characteristic tailed and skewed distribution. If we let the quantity $\phi$ be the product of a large number of iid variables, then the logarithm of $\phi$ is a sum of iid quantities and, according to the CLT, $\log \phi$ follows a normal distribution, i.e. $\phi$ follows a log-normal distribution \citep[see e.g. the simulations of][with only three multiplicative variables]{1538-4357-570-1-L21}. 

The RMS-flux relation can be explained as a consequence of a log-normal distribution of the flux \citep[][]{UttMcVa}, but the reciprocal is not true. Indeed, let the observed flux $\phi$ be a function of an underlying random variable $x$, so that $\phi = f(x)$; for a log-normal distribution, $f$ is the exponential function and $x$ is normally distributed. A small fluctuation of $x$ around $x_0$, $\delta x$, results in a small fluctuation of the flux $\phi$ around $f(x_0)$, $\delta \phi$, and the variance of the flux is
\begin{equation}
\label{Eq:VAR}
\delta \phi^2 = \left[ {{\partial f} \over {\partial x}}(x_0) \right]^2 \delta x^2.
\end{equation}

With a sample flux proportional to $f(x_0)$, a linear flux-RMS relation is equivalent to
\begin{equation}
\label{Eq:DefExp}
f(x_0)^2 \propto \left[ {{\partial f} \over {\partial x}}(x_0) \right]^2 \delta x^2.
\end{equation}

Equation~(\ref{Eq:DefExp}) is one of the definitions of the exponential function, so the sample RMS is proportional to the sample flux if and only if the flux is the exponential of an underlying variable, as is the case for a log-normally distributed flux. Log-normality thus implies linearity between the RMS and the flux\footnote{And the reciprocal is not true!}, as does any distribution that arises from the exponential of an underlying variable.

\subsection{Pareto distributions and the central limit theorem}\label{Pareto}

We now consider the class of distributions that, instead of arising from the exponential of normal distributions, arise from the exponential of exponential distributions $\exp(-\alpha x)$. These are known as Pareto distributions and are characterized by a probability density function (PDF) that follows a power-law function of index $1+\alpha$: 
\begin{equation}
f_Y(y) = {\alpha \over y^{1+\alpha}} {\rm \quad for\ } y > 1.
\end{equation}

Pareto distributions are used in seismology, in finance, and in biology and various examples of applications such as ``volcanic eruptions, solar-flares, lightning strikes, river networks, forest fires, extinctions of biological species, war casualties, Internet traffic, stock returns, insurance pay-offs'', can be found in the literature \citep[see, e.g.,][and reference therein]{2005ConPh..46..323N,Zaliapin}. 

The CLT cannot be applied to Pareto distributed iid, mostly because, for $\alpha \leq 2$, the variance of each random variable cannot be defined. A generalized central limit theorem has been established for such a high-tailed distribution \citep[see, e.g., ][]{Voit}, and it states that the sum of Pareto-distributed iid converge to $\alpha$-stable distributions, a class of probability distribution with fascinating properties.

Only a handful of $\alpha$-stable distributions can be expressed in terms of elementary functions. Some special cases are symmetric, such as the Gaussian distribution, which corresponds to the limit $\alpha = 2$, but a large skewness is generally seen, e.g., with the Landau distribution ($\alpha = 1$), used by particle physicists \citep{2000EPJC...15..163G} to describe the distribution of the energy loss of a charged particle passing through matter. In particular, for $0 < \alpha < 2$ the PDF of an $\alpha$-stable distribution asymptotically follows a power law of index $1+\alpha$ \citep{Zaliapin,nolan:2012}. 

The convergence of the CLT is achieved with a few tens of ``regular''\footnote{i.e. with finite moments.} components, but sums of Pareto iid variables tend to the asymptotic $\alpha$-stable distributions only for large numbers of components, typically $10^4$ \citep{Zaliapin}, which prevents the PDF of the sum from being analytically derived for an intermediate number of Paretian iid quantities. Even though the log-normal distribution is not an attractor for the sum of such iid variables, skewed distributions that closely resemble the log-normal can be expected, as is shown in Sect.~\ref{Sec:NBlobsFlux}.

Equations~(\ref{Eq:VAR}) and (\ref{Eq:DefExp}) prove that the sample RMS is proportional to the sample flux for a single random variable as long as it is the exponential of an underlying variable. This is the case for Pareto distributions, and it will be shown in Sect.~\ref{Sec:NBlobsFlux} that this relation holds for the sum of a large number of Paretian iid, certainly because the PDF of the sum follows the same power-law behaviour for high fluxes.

\section{A minijets-in-a-jet statistical model}\label{Sec:Model}

Power laws are generally seen as the signature of a scale invariant process and a wide variety of models can certainly produce such flux distributions. In the next sections, a simple kinematic model that naturally generates Paretian fluxes is developed.

The relativistic enhancement of the flux, known as Doppler boosting, is computed for an emitting region randomly oriented in a medium. For the sake of generality, both the emitting region and the medium, called minijet and jet respectively, have a relativistic motion. Beamed astrophysical objects thus fit in this general framework, as well as non-beamed objects, when the Lorentz factor of the medium is set to unity. 

	\subsection{Doppler factor of a minijet randomly oriented in a jet}\label{DopplBlob}

\begin{figure}[h!]
\centering
\includegraphics[width=0.81\linewidth]{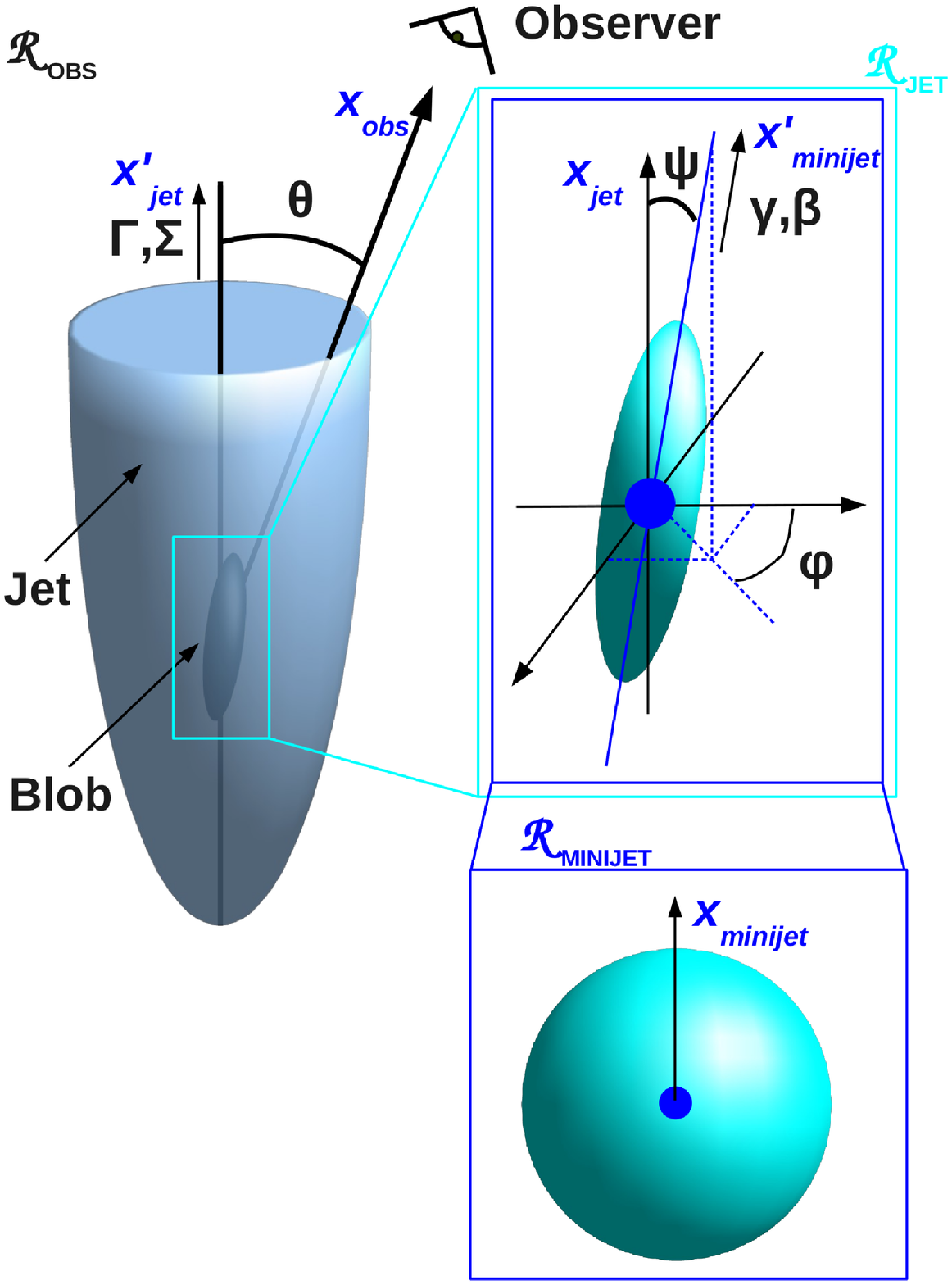}
\caption{Schematic view of the geometry. The left side corresponds to the observer frame $\cal R_{\rm obs}$, in which the jet is tilted by angle $\theta$ from the line of sight (along $x_{\rm obs}$) and is boosted by a Lorentz factor $\Gamma$ (velocity $\Sigma$) along $x'_{\rm jet}$. The minijet is defined by its Lorentz factor $\gamma$ (velocity $\beta$), in the jet frame $\cal R_{\rm jet}$ (top right). The orientation of the minijet along $x'_{\rm minijet}$ compared to the jet axis in its rest frame, $x_{\rm jet}$, is defined by the spherical angles $\psi$ and $\varphi$. The emission is assumed isotropic in the minijet frame $\cal R_{\rm minijet}$ (bottom right).}
\label{Fig:SchematicView}
\end{figure}

We consider a minijet, e.g. a blob of plasma, that moves randomly in a jet and is characterized by its Lorentz factor $\gamma$ and associated velocity $\beta = \sqrt{1-\gamma^{-2}}$, within the jet frame. The orientation of the minijet in the jet frame can be defined with the conventional angles in spherical coordinates, which will be called $\psi$ and $\varphi$. The jet is defined by a Lorentz factor $\Gamma$, associated velocity $\Sigma$, within the observer frame and its axis is misaligned from the line of sight by an angle $\theta$. A schematic representation of these parameters is shown in Fig.~\ref{Fig:SchematicView}.

To derive the Doppler factor $\delta$ of the emitting region, the minijet, one only needs to determine the energy measured in the observer frame as a function of the energy in the minijet frame, where the emission is assumed to be isotropic. We consider a photon emitted in the minijet frame with energy $E_{\rm minijet}$ and momentum $\vec p_{\rm minijet} = \{p_{x\ \rm minijet},p_{y\ \rm minijet},p_{z\ \rm minijet}\}$. A series of transformations must be applied to derive the energy in the observer frame~: the tilt of the jet, a boost, the tilt of the minijet in the jet, defined by two angles, and finally the boost of the minijet. Only photons propagating along the $x_{\rm obs}$ direction, i.e. along the line of sight, are considered because these are the only ones observed. We can thus set the transverse momentum to zero, $p_{y\ \rm obs}=p_{z\ \rm obs}=0$, while the longitudinal momentum is $p_{x\ \rm obs}~=~E$, where the speed of light is set to unity and where $E$ is the photon energy in the observer frame. 

These equalities and the frame transformations are summarized in Eq.(~\ref{Eq:AllTransformations}), where c and s denotes the cosine and sine functions:

\begin{equation}
\label{Eq:AllTransformations}
\begin{array}{l}
	\left[\begin{array}{c} E \\ E \\ 0 \\ 0 \end{array}\right] = 

	\left[\begin{array}{cccc}
	1 &  &  &  \\ 
	  & {\rm c}_\theta  & {\rm s}_\theta  &  \\
	  & -{\rm s}_\theta & {\rm c}_\theta  &  \\
	  &  &  & 1 \\
	\end{array}\right] 

	\left[\begin{array}{cccc}
	\Gamma & \Gamma \Sigma &  &  \\ 
	\Gamma \Sigma & \Gamma &  &  \\
	      &  & 1 & \\
	 & & & 1 \\
	 \end{array}\right] 

	\left[\begin{array}{cccc}
	1 &  &  &  \\ 
	  & 1  &   &   \\
	  &  & {\rm c}_{\varphi} & {\rm s}_{\varphi} \\
	  &   & -{\rm s}_{\varphi}  &  {\rm c}_{\varphi} \\
	\end{array}\right] \\ \qquad \quad
	
	\left[\begin{array}{cccc}
	1 &  &  &  \\ 
	  & {\rm c}_{\psi}  & {\rm s}_{\psi}  &  \\
	  & -{\rm s}_{\psi} & {\rm c}_{\psi}  &  \\
	  &  &  & 1 \\
	\end{array}\right]

	\left[\begin{array}{cccc}
	\gamma & \gamma \beta &  &  \\ 
	\gamma \beta & \gamma &  &  \\
	      &  & 1 & \\
	 & & & 1 \\
	\end{array}\right]  

	 \left[\begin{array}{c} E_{\rm minijet} \\ p_{x\ \rm minijet} \\ p_{y\ \rm minijet} \\ p_{z\ \rm minijet} \end{array}		\right].
	\end{array}
\end{equation}

The inverse of the Doppler factor, i.e. the ratio between the emitted and the observed energies, is derived by inverting Eq.(~\ref{Eq:AllTransformations}). The time-like component of the vectorial equality then reads as $\delta^{-1} E~=~E_{\rm minijet}$, with $\delta$ given in Eq.(~\ref{Eq:DopplerFactor}):

\begin{equation}
\label{Eq:DopplerFactor}
\delta = {1 \over { \gamma \Gamma ( 1 + \Sigma\beta{\rm c}_{\psi} - (\Sigma+\beta{\rm c}_{\psi}){\rm c}_\theta) + \gamma \beta {\rm s}_\theta {\rm s}_{\psi} {\rm c}_{\varphi} }}.
\end{equation}

For a jet closely aligned with the line of sight, the extrema of the Doppler factor are
\begin{equation}
\label{Eq:DopplerEx}
\begin{array}{c}
	\delta \leq {1 \over { \gamma \Gamma ( 1-\Sigma)(1- \beta) } } \sim 4 \Gamma \gamma \\
	\delta \geq {1 \over { \gamma \Gamma ( 1-\Sigma)(1 + \beta) } } \sim \Gamma / \gamma
\end{array}
\end{equation}
where the approximated expressions correspond to the ultra-relativistic limit.

It is worth noting that for reasonable Lorentz factors such as $\Gamma =5$ and $\gamma=5$, the maximum Doppler factor obtained for a minijet and a jet aligned with the line of sight is enhanced by a factor $2\gamma = 10$ compared to isotropic emission in the jet frame ($\gamma=1$), and Doppler factors as high as $10^2$ are reached.

\begin{figure}[h!]
\hspace{-0.25cm}
\includegraphics[width=1.05\linewidth]{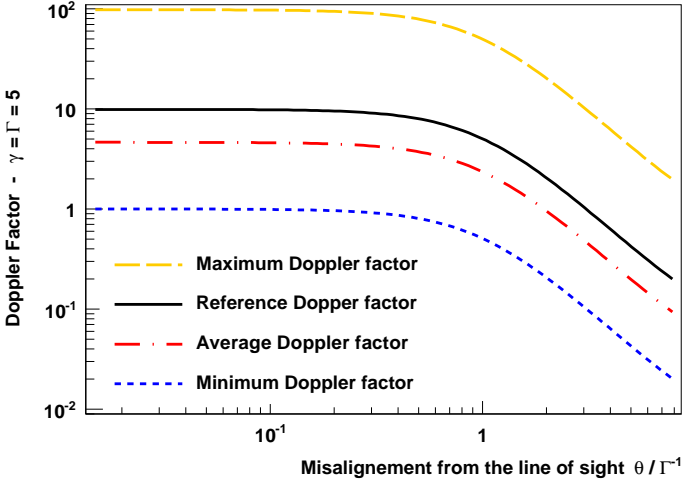}
\caption{Doppler factor of a minijet with a Lorentz factor $\gamma = 5$ moving in a jet of Lorentz factor $\Gamma = 5$ as a function of the misalignment between the jet and the line of sight $\theta$, normalized to the opening angle of the jet $1/\Gamma$. The maximum, average, and minimum Doppler factor are compared to the usual Doppler factor (solid line), derived for an isotropic emission in the jet frame ($\gamma=1$).}
\label{Fig:DopplerFactor}
\end{figure}

Figure~\ref{Fig:DopplerFactor} shows the Doppler factor as a function of the misalignment from the line of sight $\theta$. The Lorentz factors of the jet and of the minijet are fixed to a value of five and $\theta$ is normalized to $1/\Gamma$, the opening angle of the jet. 
	
The maximum and minimum values of the Doppler factor are represented with the long and short dashed lines. The average Doppler factor is derived with $\mu = \cos \psi$ uniformly distributed between [-1,1] and $\varphi$ uniformly distributed between [0,2$\pi$]. This corresponds to an isotropic distribution of the minijet orientation in the jet frame. One can compute the average Doppler factor in the blazar case, i.e for $\theta=0$
\begin{equation}
\hat \delta_0 = {1 \over {\gamma\Gamma(1-\Sigma)}} \int_{-1}^{1} \rm d\mu {1 \over {1-\beta \mu}} \nonumber
\end{equation}
\begin{equation}
\hat \delta_0 = {1 \over {\gamma\Gamma(1-\Sigma)\beta}} \ln\left({{1+\beta}\over{1-\beta}}\right) \sim {4\Gamma \over \gamma} \ln 2\gamma.
\end{equation}

For $\gamma \gtrsim 4$, the reference Doppler factor ($\gamma=1$), represented with the solid line in Fig.~\ref{Fig:DopplerFactor}, is larger than the average Doppler factor, a comparison that can be extended only to quantities that are linearly dependent on the Doppler factor. The impact on the flux intensity, which is roughly a quartic function of the Doppler factor, is discussed in the following.

\subsection{Flux of a single minijet}\label{SingleBlob}

Let $I(E)$ be the flux intensity at energy $E$, the quantity $I(E)/E^3$ is a Lorentz invariant \citep[e.g.,][]{Rybicki}, and the Doppler boost $\delta$ of the energy results in an enhancement $\delta^3$ of the intensity or of the flux \citep[see, e.g.,][]{Urry}. One finds the usual enhancement of the bolometric luminosity by a factor $\delta^4$ after integration over the energy domain.

The non-thermal spectra of high-energy astrophysical sources can generally be approximated, at least locally, by power-law functions. We characterize the intensity of the source by the index $s$ (photon index $s+1$), i.e. $I(E_{\rm minijet}) \propto E_{\rm minijet}^{-s}$, where $E_{\rm minijet}$ is the photon energy in the source rest frame. The intensity measured by the observer is, in the observer's frame,
\begin{equation}
\label{Eq:DefI}
I(E) = \delta^3 I(E_{\rm minijet}) = \delta^3 I(E/\delta) \propto \delta^3 (E/\delta)^{-s}	\propto \delta^{3+s} E^{-s}.
\end{equation}

The distribution of the intensity can then be derived from Eq.(~\ref{Eq:DopplerFactor}), imposing a distribution for the underlying random variables. We assume in this model that the relevant physical parameter is the orientation of the minijet(s) in the jet frame. Imposing isotropy in this frame corresponds to having $\psi$ and $\varphi$ angles homogeneously populate the unit sphere, i.e. $\mu~=~\cos\psi$ uniformly distributed between [-1,1] and $\varphi$ uniformly distributed between [0,$2\pi$]. For the sake of simplicity, we derive the probability density function of the intensity in the blazar case. In this case, combining the definition of the Doppler factor in Eq.(~\ref{Eq:DopplerFactor}) with Eq.(~\ref{Eq:DefI}) yields
\begin{equation}
I(E) \propto \left[\gamma\Gamma(1-\Sigma)(1-\beta\rm \mu) \right]^{-3-s} E^{-s} \equiv (4\Gamma\gamma)^{3+s} g(\mu) E^{-s}.
\end{equation}
The factor $(4\Gamma\gamma)^{3+s}$  scales the function $g$ so that
\begin{equation}
g(\mu) =  \left({{(1+\Sigma)(1+\beta)} \over 4} \times {{1-\beta} \over {1-\beta \mu}} \right)^{3+s} \leq 1 {\rm \ with\ \mu}\in[-1,1].
\end{equation}

We call $I_N = g(\mu)$ the intensity normalized to its maximum. The PDF of $I_N$, $f_\mathcal{I}(I_N)$, is linked to the PDF of $\mu=\cos \psi$, $f_{\mathcal{C}}(\mu) = 1/2$ with $\mu\in[-1;1]$ via
\begin{equation}
\label{Eq:Compos}
f_\mathcal{I}(I_N) = \left|{{\partial g^{-1}(I_N)} \over  {\partial I}}\right| f_{\mathcal{C}}\left(g^{-1}(I_N)\right),
\end{equation}
where 
$\displaystyle g^{-1}(x) = {1 \over \beta}\times\left(1-{{1+\Sigma}\over 2}\times{1\over {2\gamma^2}}\times x^{-{1 \over {3+s}}}  \right) $.

Then, the probability density function of the normalized intensity is
\begin{equation}
\label{Eq:ParDist}
f_\mathcal{I}(I_N) = {{1+\Sigma} \over 2\beta}  \times {1 \over {4\gamma^2(3+s)}} \times {1 \over {I_N^{1 + {{1} \over {3+s}}}}}.
\end{equation}

The flux emitted by a minijet randomly oriented in a jet (or in a non-relativistic medium) thus follows a Pareto distribution of index $\alpha = 1/(3+s)$, where $s$ is the spectral index of the intensity. The Pareto distribution holds for a non-beamed object ($\Gamma=1$ and $\Sigma=0$), as derived independently by \citet{2012arXiv1205.5094C} to model the flares of the Crab in the high-energy domain.

\begin{figure}[h]
\hspace{-0.25cm}
\includegraphics[width=1.02\linewidth]{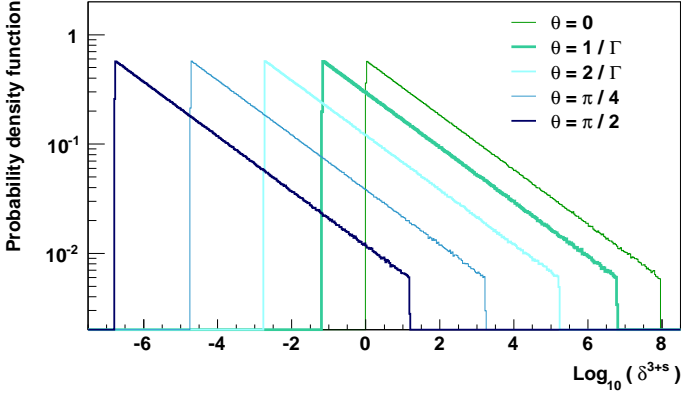}
\caption{Distribution of the logarithm of the intensity of a minijet for various angles $\theta$ between the line of sight and the jet axis. The intensity is proportional to $\delta^{3+s}$, where $\delta$ is the Doppler factor and $s$ the spectral index. These distributions are obtained simulating random orientations of a minijet for $s= 1$, for a jet boost $\Gamma = 5$, and a minijet boost $\gamma = 5$, without generality loss. While the intensity decreases when the misalignment increases, the slope of the distribution remains the same as in the ``blazar'' case, $\theta = 0$.}
\label{Fig:Distrib1blob}
\end{figure}

The Pareto distribution also holds for beamed objects misaligned from the line of sight, as shown in Fig.~\ref{Fig:Distrib1blob}, where the distribution of the logarithm of the intensity $\propto \delta^{3+s}$ has been generated by drawing random minijet orientations. The conservation of the slope of the distribution can be understood by neglecting the term $\gamma \beta {\rm s}_\theta {\rm s}_{\psi} {\rm c}_{\varphi}$ in Eq.(~\ref{Eq:DopplerFactor}), in which case the previous proof holds with the inverse of the Doppler factor remaining a linear function of $\cos \psi$. The conclusions derived in the following can then be applied to both blazars, such as \PKS, and non-aligned objects, such as M~87. 

\subsection{Flux of N minijets}\label{Sec:NBlobsFlux}

As discussed in Sect.~\ref{Pareto}, Pareto-distributed random variables do not follow the central limit theorem, and their sum does not converge to a normal distribution. Assuming that the jet is composed of several randomly oriented minijets, the total emitted flux is proportional to the sum of the $N$ independent minijet intensities, in an optically thin medium. 

To illustrate this point, minijets of Lorentz factor $\gamma = 5$ are simulated for $\theta = 0$ and a jet boost $\Gamma = 5$. These moderated values yield a single component flux spanning over eight decades, roughly between $( \Gamma / \gamma )^{3+s}$ and $( 4\Gamma \gamma )^{3+s}$. Unless strongly suppressing the beaming (typically for $\gamma \lesssim 1.5$), the dynamic range of the flux is large enough that the location of the cut-off does not affect the following results (up to at least $10^4$ additive components).

\begin{figure}[h]
\includegraphics[width=0.99\linewidth]{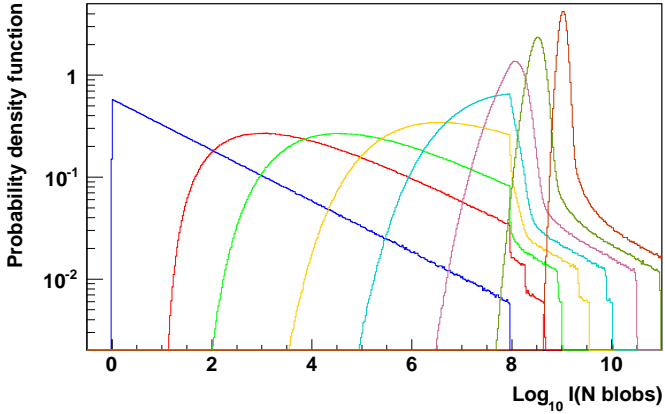}
\caption{Distribution of the logarithm of the intensity of $N$ independent and randomly oriented minijets. The number of minijets $N$ increases from left to right with $N\in\{1,10,30,10^2,3\times 10^2,10^3,3\times 10^3,10^4 \}$. Even for a large number of regions, asymmetrical, tailed distributions are obtained.}
\label{Fig:DistribNblob}
\end{figure}

As shown in Sect.~\ref{SingleBlob}, the results for other orientations or other Lorentz factors remain identical within a multiplicative factor (i.e. a shift in log representation). The flux distributions simulated for several numbers $N$ of minijets are shown in Fig.~\ref{Fig:DistribNblob}. To generate smooth distributions, $10^8$ iterations, called hereafter time steps, are performed with $N\in\{1,10,30,10^2,3\times 10^2,10^3,3\times 10^3,10^4 \}$.

\begin{figure}[h]
\includegraphics[width=0.92\linewidth]{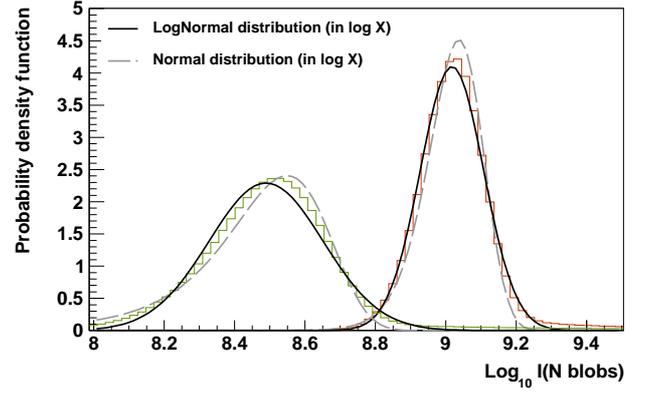}
\caption{Distribution of the logarithm of the flux of $N$ minijets for $N~=~3\times 10^3$ (left) and $N = 10^4$ (right), as in Fig.~\ref{Fig:DistribNblob}. The continuous black and grey dashed lines represent the best fit with a log-normal and normal flux distributions, respectively.}
\label{Fig:LogNormAndNormFit}
\end{figure}

For a large number of minijets, typically $N \gtrsim 10^3$, the distribution of the logarithm of the flux can be described with a peak, followed by a power-law tail. The distributions obtained for $3\times 10^3$ and $10^4$ minijets are shown Fig.~\ref{Fig:LogNormAndNormFit}, with a linear $y$-axis, together with the best-fit functions corresponding to a normal and log-normal flux.

Although the distribution of the flux of $N$ minijets is neither normal nor log-normal, Fig.~\ref{Fig:LogNormAndNormFit} illustrates how an experimental distribution, with limited statistics and a poor characterization of the high flux tail, could be interpreted as a log-normal distribution, even if arising from an additive process and not from a multiplicative one.

\begin{figure}[t!]
\centering
\includegraphics[width=0.75\linewidth]{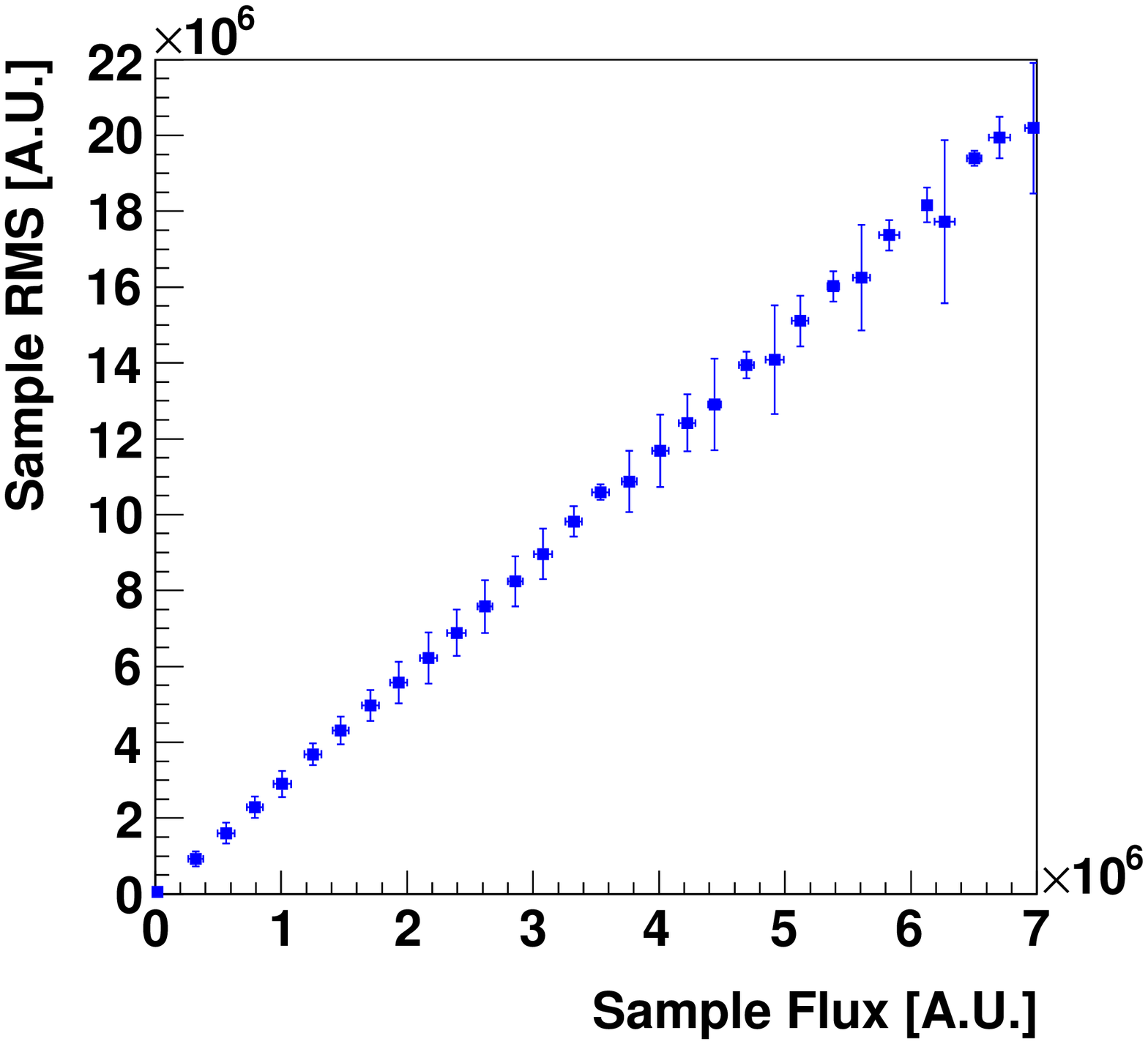}
\includegraphics[width=0.75\linewidth]{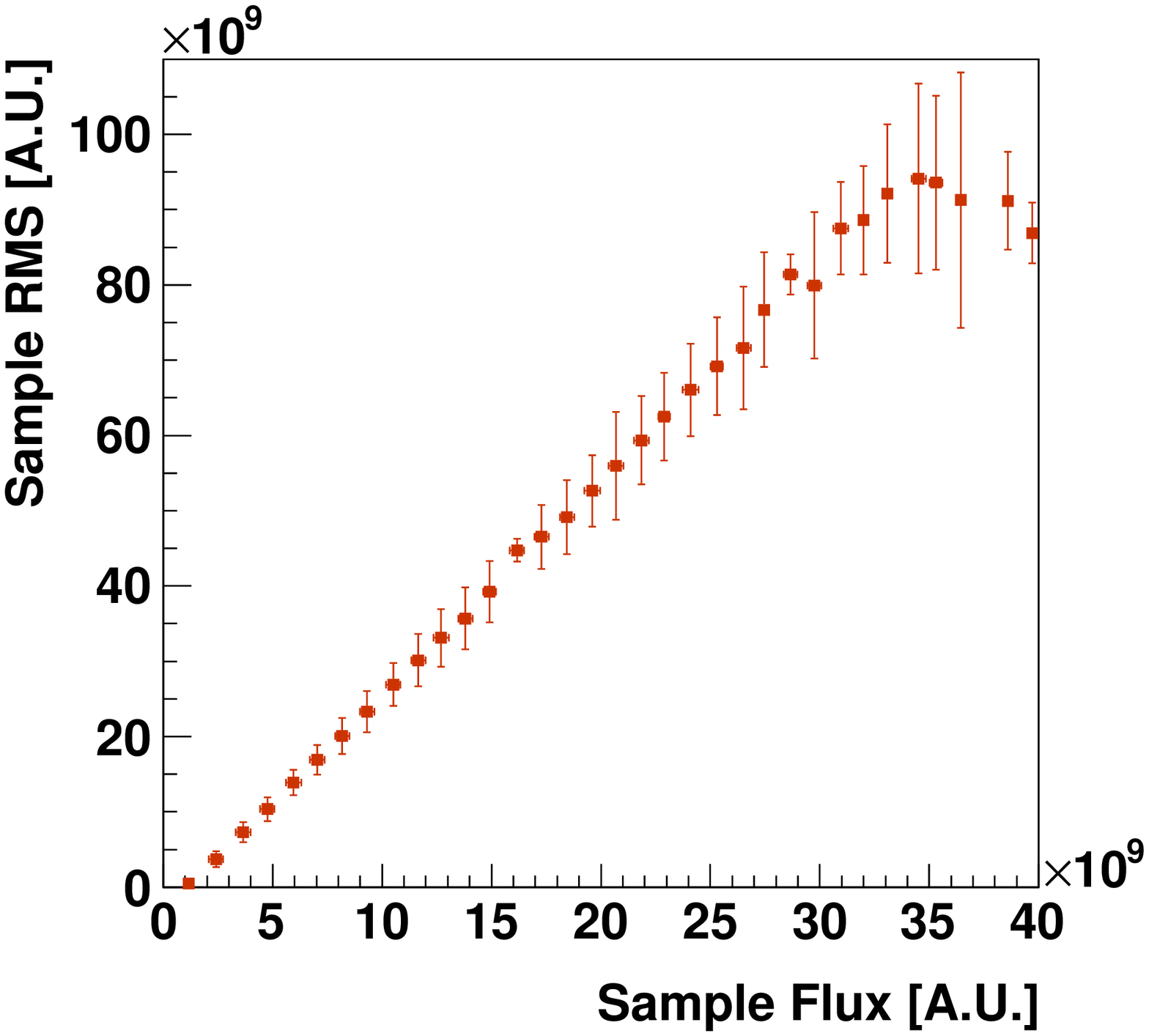}
\caption{Sample RMS as a function of the sample flux of $N=1$ minijet (top) and the sum of $N=10^4$ minijets (bottom). Linear relations are found in both cases, with a zero {\it x}-intercept in the first case and a positive one in the second.}
\label{Fig:RMSflux}
\end{figure}

We have shown in Sect.~\ref{Pareto} that the RMS is strictly proportional to the flux if and only if the flux is the exponential of an underlying variable, which is the case for Pareto distributions. Interestingly, linearity holds when considering the sum of a large number of power-law components. To illustrate this statement, light curves are simulated with $10^5$ time steps for $N = 1$ and $N = 10^4$ minijets. The sample mean and the sample RMS are then computed in ten points wide windows. For the sake of clarity, the flux and the RMS are averaged in $50$ bins between the maximum flux and the minimum flux. The sample RMS is plotted as function of the sample flux in Fig.~\ref{Fig:RMSflux}, where the error bars correspond to the standard deviation in each bin.

The positive {\it x}-intercept in the RMS-flux relation, as seen in the right-hand panel in Fig.~\ref{Fig:RMSflux}, corresponds to the peak in the flux distribution shown in Fig.~\ref{Fig:DistribNblob}. While it could be interpreted as a constant component, this threshold flux corresponds to the value around which the average emissions of the minijets pile up within the additive model presented here.

\section{Discussion and conclusion}\label{Sec:CCL}

The linear relation between the RMS and the flux of individual blazars or the skewness of the distribution of the flux are interpreted by several authors as arising from multiplicative processes, favouring a variability stemming from the disk.

We are studying a minijets-in-a-jet statistical model where the variability stems from the jet itself. We first considered the enhancement of the flux due to the relativistic Doppler effect and show that an isotropic orientation of a single region in the jet frame result in a Pareto distribution of its contribution to the flux, characterized by a power-law PDF. As for log-normal variables, Pareto-distributed variables can be seen as the exponential of an underlying variable and, as such, satisfy the linear relation between the flux and its RMS. These random variables do not fulfil the hypotheses under which the central limit theorem can be applied and the summation of the contributions of many regions does not yield a normal distribution of the flux. The total flux distribution thus remains highly skewed and tailed when increasing the number of contributing regions, up to at least $10^4$ regions, and could very well be interpreted as a log-normal. A noticeable difference between the attractor of the sum of Pareto distributed variables, called $\alpha$-stable distributions, and the log-normal distribution is their high flux tails that follow a power law and that could be probed with long-term, finely-sampled light curves. This power-law tail is, moreover, responsible for the invariance of the linear relation between the RMS and the flux when the number of minijets is increased. The study of the link between the slope of the RMS-flux relation and the kinematic parameters of the mini-jets-in-a-jet statistical model is beyond the scope of this paper and will be the object of further studies.

The minijets-in-a-jet statistical model is largely inspired from the jets-in-jet model of \citet{2009MNRAS.395L..29G}, where magnetic dissipation triggers the emission of relativistic blobs of plasma. The authors reproduce the luminosity and the fast variations observed in the light curves of blazars at very high energies with a Lorentz factor of the minijet $\gamma \sim 10$ in the jet frame. \citet{2012MNRAS.420..604N} refined this model and invoke relativistic turbulence to explain both the variability time scales and the scarcity of the flaring events. The random orientation of the minijet in the jet frame would be directly linked to the direction in which the reconnection region is created (magnetic dissipation) or to the wandering of the velocity vector of the minijet over the radiating time (relativistic turbulence). The isotropic orientation of the minijets in the jet frame is the most natural configuration in both scenarios.

A ``minijet statistical model''\footnote{motivating the name of the model we have developed.} has also recently been derived by \citet{2012arXiv1205.5094C} to explain the high-energy flares of Crab Nebula. The authors assume that magnetic reconnection triggers the emission of two-sided relativistic flows with a random orientation in the nebula frame, a particular geometry corresponding to a bulk Lorentz factor of the medium set to unity in the framework developed in this paper. Assuming that the individual components do not overlap, the authors restricted the study to a very small number of minijets, a problem that is overcome in this paper.

Unlike \citet{2012arXiv1205.5094C}, we did not study observables such as the power spectral density or the variation time scales observed in light curves. These quantities depend on the detailed mechanism responsible for the generation of the minijets, and their study is beyond the scope of this paper. Our model, based on a Doppler-boosted beamed emission, can certainly be generalized to other mechanisms, such as the ``kinetic beaming'' studied by \citet{2012arXiv1205.3210C}, which will probably yield similar conclusions on the statistical properties of the flux, while providing refined energy-dependent observables.

The generalization of such stochastic models, where many sub-regions are considered, growingly reduces the dominance of shock emission models in high-energy astrophysics. The springboard of such a change of paradigm points out the need for high-quality non-thermal light curves and widening studies of flux distributions, of the relation between the sample RMS and the sample flux, and of the power spectral density.

\acknowledgements
The authors are grateful to the anonymous referee for the insights that have improved this article.

\bibliography{MiniJetsInAJet.bbl}
\bibliographystyle{aa}

\end{document}